\RequirePackage[2020-02-02]{latexrelease}
\documentclass[aps, twocolumn, a4paper, superscriptaddress]{revtex4}

\usepackage{epsfig}
\usepackage{graphicx}
\usepackage{float}
\usepackage{amsmath, amssymb}
\usepackage{mathtools}

\allowdisplaybreaks[4]

\begin{document}

\title{Evolution of cooperation under a generalized death-birth process}

\author{Chaoqian Wang}
\email{CqWang814921147@outlook.com}
\affiliation{Department of Computational and Data Sciences, George Mason University, Fairfax, VA 22030, USA}

\author{Attila Szolnoki}
\email{szolnoki.attila@ek-cer.hu}
\affiliation{Institute of Technical Physics and Materials Science, Centre for Energy Research, P.O. Box 49, H-1525 Budapest, Hungary}

\begin{abstract}
According to the evolutionary death-birth protocol, a player is chosen randomly to die and neighbors compete for the available position proportional to their fitness. Hence, the status of the focal player is completely ignored and has no impact on the strategy update. In this work, we revisit and generalize this rule by introducing a weight factor to compare the payoff values of the focal and invading neighbors. By means of evolutionary graph theory, we analyze the model on joint transitive graphs to explore the possible consequences of the presence of a weight factor.
We find that focal weight always hinders cooperation under weak selection strength. Surprisingly, the results show a non-trivial tipping point of the weight factor where the threshold of cooperation success shifts from positive to negative infinity. Once focal weight exceeds this tipping point, cooperation becomes unreachable. Our theoretical predictions are confirmed by Monte Carlo simulations on a square lattice of different sizes. We also verify the robustness of the conclusions to arbitrary two-player prisoner's dilemmas, to dispersal graphs with arbitrary edge weights, and to interaction and dispersal graphs overlapping arbitrarily.
\end{abstract}

\maketitle

\section{INTRODUCTION}
Evolutionary game theory is a broadly used framework to understand how cooperation emerges among selfish individuals who would prefer defection individually \cite{sigmund2010calculus}. This conflict is the key obstacle when life steps onto a higher level at different stages of evolution \cite{nowak2004emergence,maynard_95,nowak2006evolutionary}. In the last decades, several mechanisms have been identified to explain this process \cite{nowak2006five}. One of them is network reciprocity which has collected significant research interest due to its broad occurrence in realistic situations \cite{szabo_pr07,perc2017statistical,roca_plr09,perc2013evolutionary}. 
To explore the possible consequences of permanent and limited interactions, evolutionary graph theory was proposed \cite{lieberman2005evolutionary,allen2014games}, and the evolution of cooperation has been studied on various graphs including
isothermal graphs \cite{allen2019evolutionary}, temporal graphs \cite{li2020evolution}, heterogeneous graphs \cite{mcavoy2020social}, multilayer graphs \cite{su2022evolutionmultilayer}, and directed graphs \cite{su2022evolutionasymmetric}. It is generally believed that structured populations often promote cooperation \cite{lieberman2005evolutionary,su2022evolutionmultilayer,nowak1992evolutionary,ohtsuki2006simple}, but not always \cite{hauert2004spatial,su2019spatial}.

The core assumption of evolutionary dynamics is that individuals tend to imitate the strategy with a higher payoff. The general sensitivity of individuals to this difference is characterized by the strength of selection. Accordingly, models can use strong \cite{nowak1992evolutionary}, intermediate \cite{szabo1998evolutionary,szabo2002phase,wang2021public,wang2022modeling,wang2022between}, or weak selection scenarios \cite{lieberman2005evolutionary}. On the one hand, lab experiments indicated an intermediate selection strength in human populations \cite{rand_pnas13,zisis_srep15}. On the other hand, one may claim that the weak selection assumption is less relevant because it almost neglects the driving force of evolution. However, the rationality behind the weak selection assumption is that various factors contribute to an individual's fitness, and the fruit of game interactions is just one of these factors \cite{ohtsuki2006simple}. Furthermore, this assumption makes calculations analytically feasible, thus becoming an attractive  playground for theoretical approaches \cite{ohtsuki_jtb06,wild_jtb07}. A key question for these calculations is to identify the threshold of ``dilemma strength'' over which cooperation is favored. For two-player games \cite{lieberman2005evolutionary}, calculations are now available for any population structure \cite{allen2017evolutionary}. 
More generally, remarkable results have also been obtained for the analytical threshold favoring cooperation in multiplayer games. For the public goods game \cite{hauser2019social}, Li {\it et al.} \cite{li2014cooperation} deduced the threshold favoring cooperation on random regular graphs, and Su {\it et al.} \cite{su2018understanding,su2019spatial} deduced the threshold favoring cooperation on transitive graphs. Some scholars also argue that it is natural to study multiplayer games on hypergraphs \cite{burgio2020evolution,alvarez2021evolutionary}, but such a perspective remains to be explored.

While the selection strength determines an individual's sensitivity to a higher payoff when considering alternative strategies, a player's willingness to change an actual strategy is another independent factor. This aspect was studied from different angles, such as strategy learning capacity \cite{szolnoki_epl07,chen_xj_ijmpc08,szolnoki_csf20b}, behavioral inertia \cite{szabo1998evolutionary,szolnoki_pre09,liu2010effects,zhang2011inertia,du2012effects,chang2018cooperation}, overconfidence \cite{johnson2011evolution,li2016coevolution,szolnoki2018reciprocity}, and stubbornness \cite{cimpeanu_srep22,szolnoki_pre14b,cimpeanu_kbs21}. Conceptually, it can be related to the self-loops of the nodes \cite{tkadlec_ncom21,tkadlec_pcbi20} (but this work does not consider self-loops at the graph level but studies focal weight independently). The mutual idea of these concepts is to introduce the focal agent's weight in the strategy updating. When this weight is high, agents have higher inertia/overconfidence; thus, they are more reluctant to change strategy. Importantly, there is a significant difference between the selection strength and the focal weight: while the consequence of selection strength could be bidirectional and improves (weakens) reproduction activity for a higher (lower) payoff, the impact of focal weight on strategy update is unidirectional by simply decreasing its probability.

At first, one might expect that introducing the same focal weight value for all individuals seems to be a strategy-neutral modification. As a result, its consequence on the competition of strategies is not apparent. Indeed, some previous works revealed that moderate focal weight in strategy updating could promote cooperation \cite{liu2010effects,du2012effects,chang2018cooperation}. These studies, however, applied numerical simulations on structured populations or theoretical analysis on well-mixed populations. In this work, we provide a theoretical analysis of a finite structured population by utilizing the so-called identity-by-descent (IBD) method of evolutionary graph theory \cite{allen2014games,su2019spatial}.

Our principal goal is to analytically explore the impact of the focal weight concept on the evolution of cooperation in structured populations. From this viewpoint to use the so-called death-birth strategy update is a logical choice because, traditionally, this protocol completely ignores the status of the focal player, which can be considered as a zero-weight limit. In the generalized case, by introducing a nonzero weight, we can gradually leave the classic dynamics and reveal the consequences of the modified dynamical rule. Technically, we apply the similar concept introduced by Su {\it et al.} \cite{su2019spatial} who considered the focal weight concept during the interactions. In our case, however, the focal weight determines the strategy update probability, not the payoff values originating from interactions. In the following, we define our model where the introduction of focal weight can be considered as an extension of the classic death-birth dynamical rule. 

\section{MODEL}\label{def}
\subsection{Joint transitive graphs and random walks}\label{graphs}
The population structure can be described by an interaction graph $\mathcal{G}_I$ and a dispersal graph $\mathcal{G}_R$. They are joint, which means they share the same node set $V=\{1,2,\dots,N\}$. Each node represents a player, where the population size is $N$. The joint interaction and dispersal graphs are both transitive: for all nodes $i$ and $j$, there is an isomorphism that transforms $i$ into $j$ \cite{su2019spatial,taylor2007evolution,debarre2014social}. Intuitively speaking, the nodes are uniform in perceiving the whole network structure. Common transitive graphs include but are not limited to ring networks, lattices with periodic boundary conditions, and fully connected populations.

Players play the games on the interaction graph $\mathcal{G}_I$ and update strategies on the dispersal graph $\mathcal{G}_R$. This work focuses on strategy updating; hence we simplify the interaction graph by assuming it is unweighted. Due to transitiveness, each node has the same degree; hence they all have $k$ neighbors on $\mathcal{G}_I$. Given a node, each link on the interaction graph has the same weight of $1/k$.
On the contrary, we assume a weighted dispersal graph. For a node $i$, each link to a neighbor $j$ on $\mathcal{G}_R$ may have a different weight, denoted by $e_{ij}$, yielding $\sum_{j\in V}e_{ij}=1, \forall i\in V$. In addition, we assume symmetry ($e_{ij}=e_{ji}, \forall i,j\in V$, i.e., all graphs are undirected in this work) and self-loop is excluded ($e_{ii}=0, \forall i\in V$) both for the dispersal graph and the unweighted interaction graph.

To quantify the payoff values obtained from the game, we define $(n,m)$-random walk on the joint graphs with $n$ steps on $\mathcal{G}_I$ and $m$ steps on $\mathcal{G}_R$ (no sequential requirement) \cite{su2019spatial,debarre2014social}. The probability that an $(n,m)$-random walk ends at the starting node is denoted by $p^{(n,m)}$. The probability that an $(n,m)$-random walk ends at a cooperative player is denoted by $s^{(n,m)}$. The expected payoff of players where an $(n,m)$-random walk ends is denoted by $\pi^{(n,m)}$. Because of transitivity, we can use the same notations of $p^{(n,m)}$, $s^{(n,m)}$, and $\pi^{(n,m)}$ for all nodes over stationary distribution. Next, we define the applied game which determines the payoff values of players.

\subsection{Playing games on the interaction graph}
According to the evolutionary protocol, we randomly select a player to update its strategy.
The selected focal player plays $k$ games with its neighbors on the interaction graph. This work focuses on the simplest two-player game, the donation game \cite{ohtsuki2006simple}. In each donation game, players can adopt one of the two strategies: cooperation ($C$) or defection ($D$). Cooperation means donating $c$ to the recipient, and the other player receives an enlarged benefit $b$ ($b>c$). Alternatively, a defector player donates nothing to the partner. The total payoff of a player is the average over the $k$ games with its neighbors.

By using the terminology of random walks, we can write the expected payoff of the focal player where an $(n,m)$-random walk ends as 
\begin{equation}\label{pitwoplayer}
	\pi^{(n,m)}=-cs^{(n,m)}+bs^{(n+1,m)}.
\end{equation}
Here the first term is the donation of the focal player while the second term is the benefit originating from neighbors on the interaction graph.

\subsection{Updating strategies on the dispersal graph}
Having determined the payoff values of involved players, the focal player updates its strategy by the generalized death-birth rule with the consideration of fitness. We assume the fitness $F_i$ of player $i$ is calculated by $F_i=1-\delta+\delta\pi_i$ \cite{su2019spatial,su2018understanding}, where $\pi_i$ is the payoff of player $i$ and $\delta$ is the strength of selection. This work assumes weak selection in the $\delta\to 0$ limit. 

Importantly, we propose a new parameter, $w$ ($0\leq w<1$), to measure the focal player's weight in the strategy updating process. When comparing fitness, the focal player measures its own fitness with weight $w$, and the fitness of other players with weight $1-w$. Intuitively, a greater $w$ implies less motivation to change the strategy. 

According to the extended dynamical rule, the strategy updating probability depends not only on the neighbors' fitness, but also on the fitness of the focal player via appropriate weight factors. More precisely, the focal player $i$ copies the strategy of neighboring player $j$ with a probability
\begin{equation}\label{groupupdate}
	P(i\gets j)=\frac{(1-w)e_{ij}F_j}{wF_i+(1-w)\sum_{l\in V}e_{il}F_l},~~~~\mbox{for }j\in V.
\end{equation}
Otherwise, player $i$ does not change its strategy. As we argued previously, Eq.~(\ref{groupupdate}) gives back the classic death-birth rule in the $w=0$ limit where the state of focal player $i$ has no role. In the other $w=1$ limit, the focal player keeps its original strategy and the system remains trapped in the initial state. Between these extreme cases, we can explore how a non-zero weight value (i.e., a certain unwillingness of players to change strategies) may influence the evolution of cooperation in a structured population when $0<w<1$.

\section{Theoretical analysis}
\subsection{The general condition of cooperation success}\label{seccondi}
In the following, we employ the previously mentioned IBD method to determine the necessary condition for successful cooperator spread \cite{nowak2010evolution,allen2014games}. In the low mutation limit $\mu\to 0$, where $\mu$ denotes the mutation rate, the condition favoring cooperation over defection has the following form \cite{nowak2010evolution}: 
\begin{equation}\label{bd}
	\left\langle\frac{\partial}{\partial\delta}(\mathcal{B}_i-\mathcal{D}_i)\right\rangle_{\begin{smallmatrix}\delta=0\\s_i=C\end{smallmatrix}}>0,
\end{equation}
where the focal player $i$ is the only initial cooperative player in the system. Here, $\mathcal{B}_i$ denotes the probability that player $i$ reproduces its strategy, and $\mathcal{D}_i$ denotes the probability that player $i$ is replaced. Moreover, $\left\langle\cdot\right\rangle_{\begin{smallmatrix}\delta=0\\s_i=C\end{smallmatrix}}$ means the average over the stationary distribution under neutral drift with a single cooperator player $i$.

The main goal of our theoretical analysis is to provide an analytical threshold for cooperation success under the generalized death-birth rule. This can be done by calculating the condition~(\ref{bd}). Evidently, a lower threshold means an easier condition for cooperation to spread.

In addition, we utilize the low mutation expansion taken from Ref.~\cite{allen2014games}, which is generally valid on transitive graphs. Namely,
\begin{equation}\label{equa}
	s^{(n,m)}-s^{(n,m+1)}=\frac{\mu}{2}(Np^{(n,m)}-1)+\mathcal{O}(\mu^2),
\end{equation}
where the last term $\mathcal{O}(\mu^2)$ can be neglected.

\subsection{Application to the generalized death-birth rule}
As we stressed, in the generalized death-birth rule, we consider the status of the focal player via a weight factor, which has importance when directly calculating the condition~(\ref{bd}). But first, we need to clarify the following terms.

The ``Death" of player $i$ can be described as follows. When choosing the focal player, player $i$ is selected with a probability $1/N$. After, player $i$ adopts the strategy of a neighboring player $j$ with the probability given by Eq.~(\ref{groupupdate}). 
Alternatively, the ``Birth" process of player $i$ is the following. To be the focal player, a neighbor $j$ of player $i$ is selected with probability $1/N$. Then, the focal player $j$ adopts the strategy of player $i$ with the probability given by Eq.~(\ref{groupupdate}).
Therefore, $\mathcal{D}_i$ and $\mathcal{B}_i$ can be written as
\begin{subequations}\label{bdgroup}
	\begin{align}
		\mathcal{D}_i&=\frac{1}{N}\frac{\sum_{j\in V}(1-w)e_{ij}F_l}{wF_i+(1-w)\sum_{l\in V}e_{il}F_l}, \\
		\mathcal{B}_i&=\frac{1}{N}\sum_{j\in V}\frac{(1-w)e_{ji}F_i}{wF_j+(1-w)\sum_{l\in V}e_{jl}F_l}.
	\end{align}
\end{subequations}

By using these two terms and considering that the fitness is $F=1-\delta+\delta\pi$, the requested condition~(\ref{bd}) can be calculated as follows:
\begin{align}\label{condigroup}
	&\left\langle\frac{\partial}{\partial\delta}(\mathcal{B}_i-\mathcal{D}_i)\right\rangle_{\begin{smallmatrix}\delta=0\\s_i=C\end{smallmatrix}}>0\nonumber
	\\
	\Leftrightarrow&~\frac{1-w}{N}\left(
	\left\langle\pi_i\right\rangle_{\begin{smallmatrix}\delta=0\\s_i=C\end{smallmatrix}}
	-w\left\langle \sum_{j\in V}e_{ji}\pi_j\right\rangle_{\begin{smallmatrix}\delta=0\\s_i=C\end{smallmatrix}}\right.
	\nonumber\\
	&\left.-(1-w)\left\langle \sum_{j\in V}e_{ji}\sum_{l\in V}e_{jl}\pi_l\right\rangle_{\begin{smallmatrix}\delta=0\\s_i=C\end{smallmatrix}}
	\right)\nonumber
	\\
	&-\frac{1-w}{N}\left(-w
	\left\langle\pi_i\right\rangle_{\begin{smallmatrix}\delta=0\\s_i=C\end{smallmatrix}}
	+w\left\langle \sum_{l\in V}e_{il}\pi_l\right\rangle_{\begin{smallmatrix}\delta=0\\s_i=C\end{smallmatrix}}
	\right)>0\nonumber
	\\
	\Leftrightarrow&\left\langle\pi_i\right\rangle_{\begin{smallmatrix}\delta=0\\s_i=C\end{smallmatrix}}
	-\frac{2w}{1+w}\left\langle \sum_{j\in V}e_{ij}\pi_j\right\rangle_{\begin{smallmatrix}\delta=0\\s_i=C\end{smallmatrix}}\nonumber\\
	&-\frac{1-w}{1+w}\left\langle \sum_{j,l\in V}e_{ji}e_{jl}\pi_l\right\rangle_{\begin{smallmatrix}\delta=0\\s_i=C\end{smallmatrix}}>0.
\end{align}

Player $i$ being the starting node of the random walk, Eq.~(\ref{condigroup}) can be written as
\begin{equation}\label{walkcondigroup}
	\pi^{(0,0)}-\frac{2w}{1+w}\pi^{(0,1)}-\frac{1-w}{1+w}\pi^{(0,2)}>0,
\end{equation}
which is a specific form of the condition~(\ref{bd}). To get an explicit form for the threshold value, we first need to transform the expression of Eq.~(\ref{equa}) in the following way:
\begin{align}\label{equagroup}
	&~s^{(n,m)}-\frac{2w}{1+w}s^{(n,m+1)}-\frac{1-w}{1+w}s^{(n,m+2)}\nonumber\\
	=&~\frac{2w}{1+w}\left(s^{(n,m)}-s^{(n,m+1)}\right)\nonumber\\
	&+\frac{1-w}{1+w}\left(s^{(n,m)}-s^{(n,m+1)}+s^{(n,m+1)}-s^{(n,m+2)}\right)\nonumber\\
	=&~\frac{\mu}{2}\left(Np^{(n,m)}+\frac{1-w}{1+w}Np^{(n,m+1)}-\frac{2}{1+w}\right)\nonumber\\
	&+\frac{2}{1+w}\mathcal{O}(\mu^2),
\end{align}
where the last term proportional to $\mathcal{O}(\mu^2)$ can be neglected. In the following, we utilize the simplified payoff structure of the donation game.

\subsection{Theoretical threshold for donation game}\label{secdgtheo}
To obtain the requested threshold value for the donation game, we start from Eq.~(\ref{walkcondigroup}), transform $\pi^{(n,m)}$ to $s^{(n,m)}$ by using Eq.~(\ref{pitwoplayer}), and substitute $s^{(n,m)}$ with $p^{(n,m)}$ by using Eq.~(\ref{equagroup}). That is,
\begin{align}\label{calcugrouptwo}
	&~\pi^{(0,0)}-\frac{2w}{1+w}\pi^{(0,1)}-\frac{1-w}{1+w}\pi^{(0,2)}>0\nonumber
	\\
	\Leftrightarrow&
	~\left(-cs^{(0,0)}+bs^{(1,0)}\right)-\frac{2w}{1+w}\left(-cs^{(0,1)}+bs^{(1,1)}\right)\nonumber\\
	&-\frac{1-w}{1+w}\left(-cs^{(0,2)}+bs^{(1,2)}\right)>0\nonumber
	\\
	\Leftrightarrow&
	-c\left(s^{(0,0)}-\frac{2w}{1+w}s^{(0,1)}-\frac{1-w}{1+w}s^{(0,2)}\right)\nonumber\\
	&+b\left(s^{(1,0)}-\frac{2w}{1+w}s^{(1,1)}-\frac{1-w}{1+w}s^{(1,2)}\right)>0\nonumber
	\\
	\Leftrightarrow&
	-c\left(Np^{(0,0)}+\frac{1-w}{1+w}Np^{(0,1)}-\frac{2}{1+w}\right)\nonumber\\
	&+b\left(Np^{(1,0)}+\frac{1-w}{1+w}Np^{(1,1)}-\frac{2}{1+w}\right)>0.
\end{align}

It is easy to see that $p^{(0,0)}=1$, because one stays at the original position in the absence of movement. Similarly, $p^{(1,0)}=p^{(0,1)}=0$, because self-loop is not allowed, one cannot leave from and return to the initial node within a single step. 
The calculation of $p^{(1,1)}$, however, is case-dependent. The general case where the two graphs overlap in an arbitrary way will be discussed in Sec.~\ref{extension}. Here, we consider a common situation when a node shares the same neighbors on the interaction and dispersal graphs. To characterize the local structure we calculate the so-called Simpson degree \cite{allen2013spatial}. As we noted, there are $k$ neighbors to choose from on the interaction graph, and all of them are chosen with the same $1/k$ probability. In our present case, these neighbors are also neighbors on the dispersal graph. The probability of a step between node $i$ and a neighboring $l$ is $e_{li}$. Therefore, $p^{(1,1)}=\sum_{l\in V}{1/k\times e_{li}}=1/k$. By these $p^{(n,m)}$ values, we can calculate the threshold value,  yielding:
\begin{equation}\label{pointgrouptwo}
	\frac{b}{c}>\frac{N-2+Nw}{N-2k-Nw}k\equiv
	\left(\frac{b}{c}\right)^*,
\end{equation}
where the expression on the right-hand side of ``$>$'' is denoted by $(b/c)^*$, and ``$>$" holds if $(b/c)^*>0$. Here, the value of $(b/c)^*$ identifies the threshold over which cooperation is favored. When $(b/c)^*<0$, the result of Eq.~(\ref{pointgrouptwo}) should be $b/c<(b/c)^*$, which means cooperation is unreachable because $b/c>0$ always holds.

\begin{figure}
\centering
\includegraphics
[width=0.47\textwidth]
{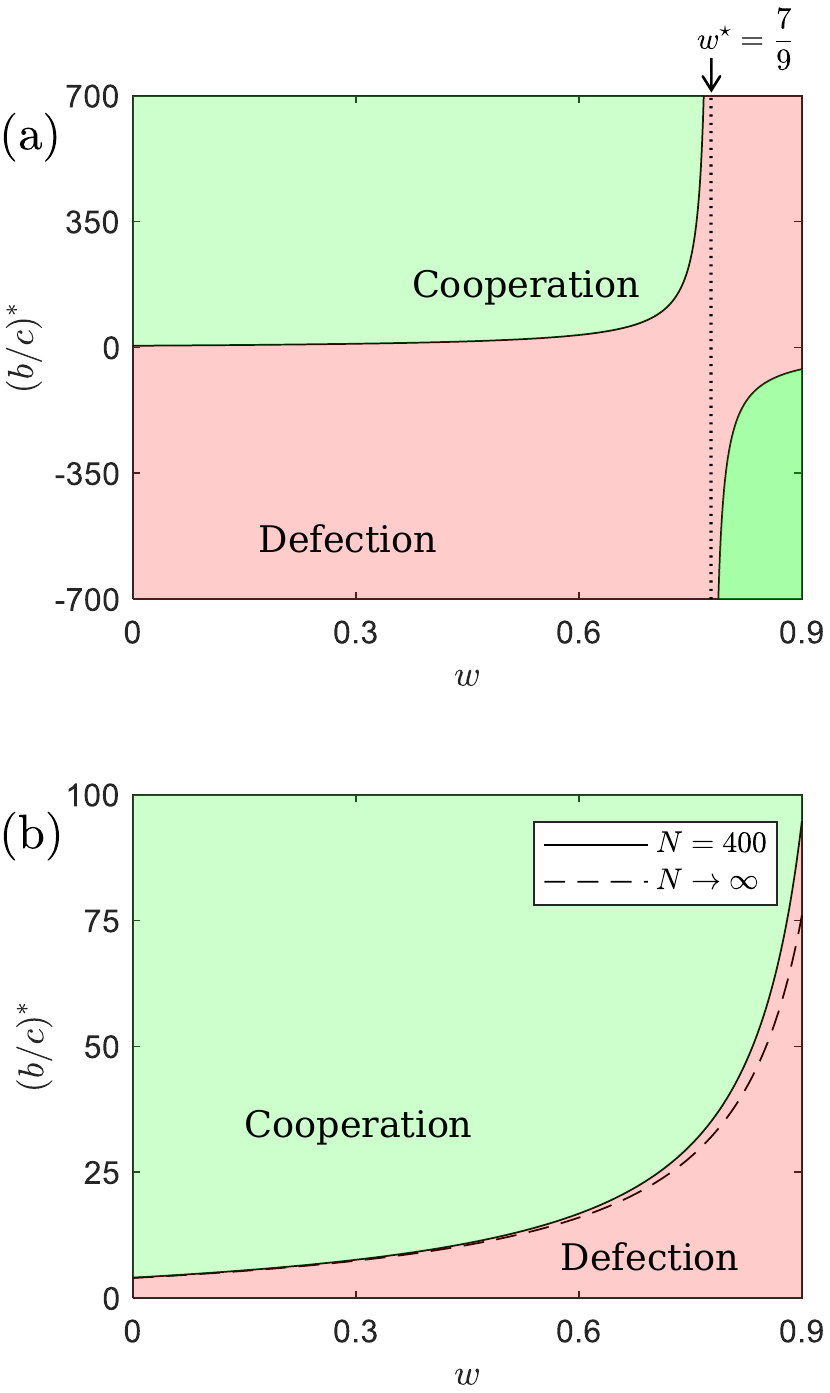}
\caption{The $(b/c)^*$ threshold for the success of cooperation as a function of $w$ described by Eq.~(\ref{pointgrouptwo}) with $k=4$. Panel~(a) shows a system of $N=36$ players. The threshold has a tipping point at $w^\star=1-2k/N=7/9$. If $w<w^\star$, then $(b/c)^*>0$, and cooperation is favored when $b/c>(b/c)^*$. If $w>w^\star$, then $(b/c)^*<0$; therefore, cooperation can never be reached. Panel~(b) shows the cases of $N=400$ and $N\to \infty$. When $N=400$, we have $w^\star=0.98$. In the range of $w<0.9$, cooperation is favored if $b/c>(b/c)^*$. The curves are close to each other, signaling that a population of $N=400$ players represents a sufficiently large system size for the approximation.}
\label{figtwoanaly}
\end{figure}

To give a deeper insight into how the threshold values depend on the weight factor, we consider a specific topology of square lattices with the von~Neumann neighborhood ($k=4$). The $w$ dependence of $(b/c)^*$ is shown in Fig.~\ref{figtwoanaly}, as calculated by Eq.~\eqref{pointgrouptwo}. Panel~(a) depicts the case where the population size is $N=36$. Here, we can detect a tipping point in $(b/c)^*$ at $w=w^\star$. When $w<w^\star$, the evolution favors cooperation if $b/c>(b/c)^*$. As panel (a) shows, the threshold $(b/c)^*$ increases by increasing $w$, which means favoring cooperation becomes more difficult for a larger focal weight. When $w>w^\star$, cooperation would be favored if $b/c<(b/c)^*$, which means cooperation is unreachable due to the $b>c>0$ constraint of the donation game. At these parameters, the tipping point is at $w^\star=7/9$, marked by a vertical dotted line in Fig.~\ref{figtwoanaly}(a). 

The position of the $w^\star$ tipping point, where the value of $(b/c)^*$ flips from positive to negative infinity, can be given by the following form:
\begin{equation}\label{tipping}
	w^\star=1-\frac{2k}{N}.
\end{equation}
This formula suggests that $w^\star<1$ always holds when $k<N/2$. In this case, there is always a tipping point if cooperation is favored under the classic death-birth process (i.e., $(b/c)^*>0$ at $w=0$). Otherwise, when $k>N/2$, we have $w^\star<0$ and cooperation is unreachable, both for the traditional ($w=0$) and the generalized ($w>0$) updating rules.

Next, we apply a sufficiently large system size $N=400$, as depicted in Fig.~\ref{figtwoanaly}(b). Here the tipping point is at $w^\star=0.98$, which is very close to the $w=1$ limit case. It practically means that cooperation can be reached for almost all $w$ values. The solid line of this system size shows that the $(b/c)^*$ threshold value increases monotonously by increasing $w$. Therefore, increasing the focal weight, or strengthening players' willingness to keep their original strategies, makes cooperation harder. 

For comparison, we also present the threshold values for the $N\to \infty$ limit. The analytical formula of the $(b/c)^*$ threshold value for $N\to \infty$ can be written as
\begin{equation}\label{largepointgrouptwo}
	\left(\frac{b}{c}\right)^*_{N\to\infty}=\frac{1+w}{1-w}k,
\end{equation}
which is a generalization of the well-known $b/c>k$ rule \cite{ohtsuki2006simple}. The dashed line in Fig.~\ref{figtwoanaly}(b) shows this function. We can see that the curves of $N=400$ and $N\to \infty$ are close to each other for almost all $w$ values, indicating that $N=400$ is sufficient (especially when $w$ is small) to be considered as a large population for the weak-selection limit, especially when $w$ is small. For example, the difference is less than 5\% when $w<0.6$.

\section{Numerical simulation}\label{numeric}
To support our analytical predictions, we present the results of Monte Carlo (MC) simulations also using $L \times L$ square lattice topology with periodic boundary conditions. Accordingly, the total size of the population contains $N=L^2$ players. Following the standard simulation protocol, we randomly assign each agent's strategy by cooperation or defection, which provides $\rho_C\approx 0.5$ portion of cooperative agents when we launch the evolution. 

During an elementary step, we randomly select a focal agent who plays the donation game with the four nearest neighbors. The average of the resulting payoff values determines the fitness of this agent according to the weak selection approach. The payoff and fitness values of neighbors are calculated in the same manner.  After, the focal player adopts the strategy of a neighbor with the probability determined by the extended death-birth rule of Eq.~\eqref{groupupdate}. For a full MC step, we repeat the above-described procedure $N$ times, which ensures that each agent is selected once on average. 

\begin{figure}
	\centering
	\includegraphics
	[width=0.48\textwidth]
	{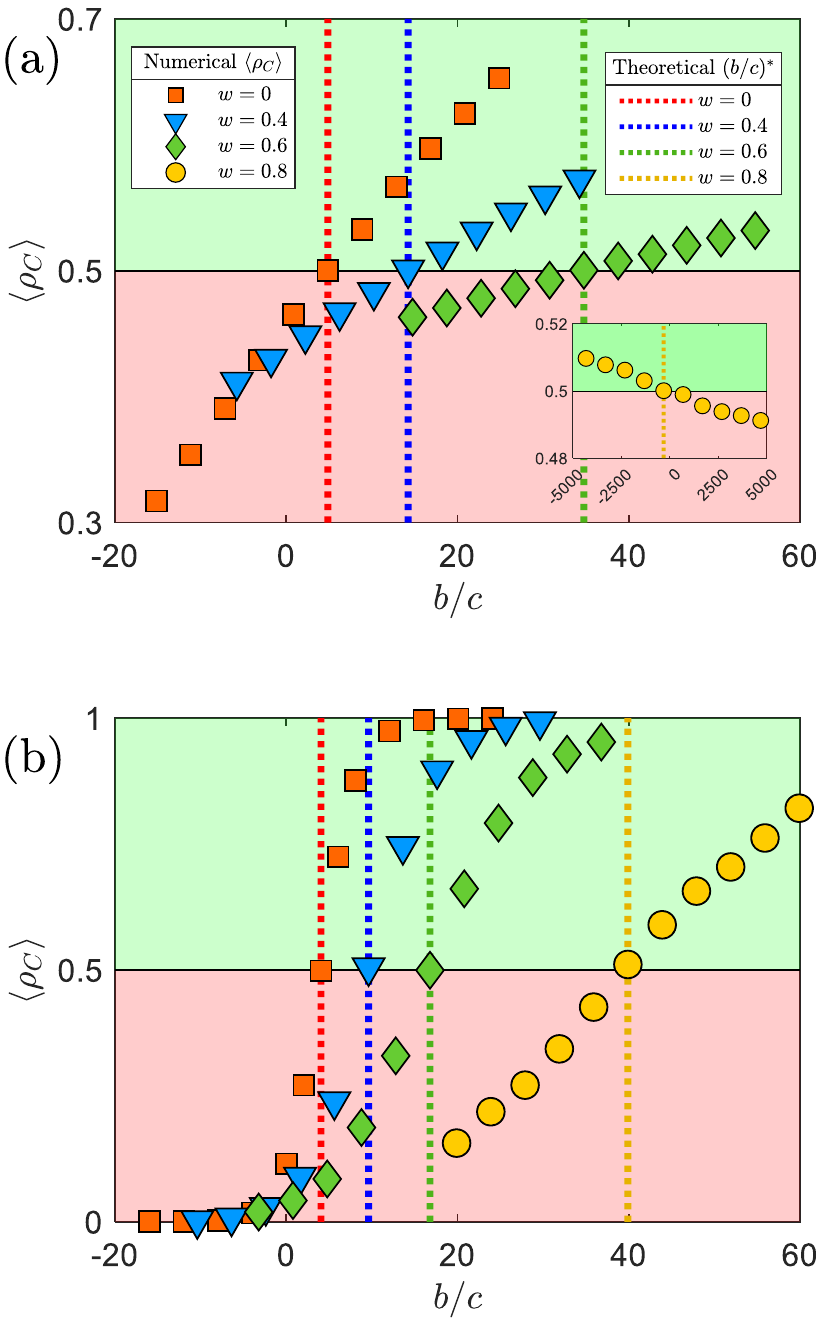}\\
	\caption{MC simulations on square lattices of different sizes where the $b/c$ control parameter is varied by keeping $c=1$ fixed. The fractions of cooperators are plotted for different focal weight values, as indicated in the legend. Panel~(a) shows the results of $L=6$ linear size, where we average over $10^6$ independent runs. Panel~(b) depicts the results obtained for $L=20$, where we average over $10^4$ independent runs. In both panels we use $\delta=0.01$ selection strength, but the inset of panel~(a) shows results obtained for $\delta=0.0001$. Dashed vertical lines represent the position of theoretical threshold level for cooperation success. The numerical results confirm the analytical predictions for all weight values.}\label{figtwonume}
\end{figure}

An independent run contains up to $4\times 10^5$ full MC steps. This relaxation time can be considered sufficiently long \cite{allen2017evolutionary} because, under a weak selection strength, the system can easily reach full cooperation ($\rho_C=1$) or full defection ($\rho_C=0$) absorbing states. If the system does not fix within the period mentioned above, then we take the portion of cooperative agents $\rho_C$ at the last time step as the result. To get reliable statistics, we perform independent runs $10^4-10^6$ times (depending on the system size) and average them. The resulting $\langle \rho_C\rangle$ values obtained for different weight factors and selection strengths are summarized in Fig.~\ref{figtwonume}.

Figure~\ref{figtwonume}(a) shows the results obtained for a $6\times 6$ lattice, where we applied $w=0$, $0.4$, $0.6$, and $0.8$ weight factors. The benefit to cost portion is varied by increasing $b$ while $c=1$ remained fixed. As expected, by increasing $b/c$, the portion of cooperators grows for small weight factors.  The critical $(b/c)^*$ value is identified where $\langle \rho_C\rangle$ exceeds 0.5. For comparison, we also mark by vertical dashed lines the positions of $(b/c)^*$ threshold values obtained from Eq.~\eqref{pointgrouptwo}. These values are $(b/c)^*=4.85$, $14.23$, and $34.75$ for $w=0$, $0.4$, and $0.6$, respectively. For completeness, we also study the $w=0.8$ case, which is beyond the $w^\star$ tipping point for this system size. Our theory predicts $(b/c)^*=-314$ here. Indeed, as the inset of panel~(a) illustrates, $\langle \rho_C\rangle$ decreases by increasing $b/c$, and the evolution favors cooperation only when $b/c<(b/c)^*$. We can see that the cooperation-favored parameter areas are consistent with those shown in Fig.~\ref{figtwoanaly}(a) for the same parameter values.

According to our theory, a $20\times 20$ system size can be considered comparable to the large population limit where the tipping point is very close to $w=1$ and the system behavior is qualitatively similar for all the mentioned $w$ values. To check this, we also present results obtained for this system size. Our observations, shown in Fig.~\ref{figtwonume}(b), confirm that the system behaves similarly for all studied $w$ values and $\rho_C$ always increases as we increase $b/c$. The theoretical threshold values are $(b/c)^*=4.06$, $9.62$, $16.79$, and $39.89$, for $w=0$, $0.4$, $0.6$, and for $w=0.8$, respectively. Vertical dashed lines mark these values, which are consistent with the $b/c$ values where $\langle \rho_C\rangle$ exceeds 0.5 in our numerical simulations. 

The reason why we can link the theoretical $(b/c)^*$ value to the location where $\rho_C$ exceeds 0.5 is the following. It is a well-known result that in the $\delta=0$ limit under neutral drift the system eventually terminates in one of the homogeneous absorbing states, and the probability of reaching the full cooperation state depends on the initial $N_C/N$ portion of cooperators \cite{cox_ap83,cox_ap86,nowak2004emergence}. For example, in our theoretical analysis, we analyze the case of starting with $N_C=1$ cooperative agent, which means the system achieves full cooperation with probability $1/N$ when $\delta=0$; hence, the sign of cooperation success is when $\langle \rho_C\rangle>1/N$. Meanwhile, the system's property (whether favoring cooperation) is independent of the initial state. Therefore, once we deduce the condition of cooperation success, such a condition indicates the probability of starting with $N_C$ cooperators ending in full cooperation $\langle \rho_C\rangle>N_C/N$ \cite{chen2013sharp}. In our numerical simulation, we initially assign each agent's strategy by random and $N_C/N\approx 0.5$; therefore, $\langle \rho_C\rangle>0.5$ is the direct sign of the evolution favoring cooperation.

\section{Extension to different graphs and games}\label{extension}
Until this point, we assumed that the interaction and dispersal graphs overlap where square lattices with the von Neumann neighborhood provided a testable topology. In the following, we leave this strong restriction to check the robustness of our observations and to see how the focal weight changes the system behavior. Moreover, we also consider the model in alternative games.

First, we show the robustness to different graphs. We still keep the basic assumptions: the interaction graph is unweighted, the self-loop is excluded, and the joint graphs are transitive. In this way, the calculation $p^{(0,0)}=1$, $p^{(1,0)}=0$, $p^{(0,1)}=0$ is invariant. To determine $p^{(1,1)}$, however, remains an open task. Therefore, the general solution of Eq.~(\ref{calcugrouptwo}) is
\begin{equation}\label{pointgrouptwogeneral}
	\left(\frac{b}{c}\right)^*=\frac{(1+w)N-2}{(1-w)Np^{(1,1)}-2},
\end{equation}
where $0\leq p^{(1,1)}\leq 1$ because $p^{(1,1)}$ denotes a probability. The general form of the tipping point $w^\star$ is
\begin{equation}\label{tippinggeneral}
	w^\star=1-\frac{2}{Np^{(1,1)}}.
\end{equation}

From Eq.~(\ref{pointgrouptwogeneral}) and Eq.~(\ref{tippinggeneral}), the tipping point $w^\star$ exists between 0 and 1 when $p^{(1,1)}>2/N$. In this case, the threshold $(b/c)^*$ increases with $w$ in $0\leq w< w^\star$ and flips from positive to negative infinity at $w=w^\star$. When $p^{(1,1)}<2/N$, we have $w^\star<0$, and cooperation is never favored.
In sum, the statement holds given $0\leq p^{(1,1)}\leq 1$, which verifies the robustness of the conclusions to dispersal graphs with arbitrary edge weights. The robustness also holds on interaction and dispersal graphs overlapping in arbitrary ways.

Second, we show the robustness to different games. We investigate the conclusions in arbitrary two-player prisoner's dilemmas, depicted by four parameters $R$, $S$, $T$, and $P$, where $T>R>P>S$. In agreement with the general notation, the payoff of a cooperative player is $R$ if the other player cooperates and $S$ if the other player defects. Also, the payoff of a defective player is $T$ if the other player cooperates and $P$ if the other player also defects. In particular, we have $R=b-c$, $S=-c$, $T=b$, $P=0$ for the donation game.

According to the Structure Coefficient Theorem proposed by Tarnita {\it et al.} \cite{tarnita2009strategy}, the condition of evolution favoring cooperation is $\sigma R+S>T+\sigma P$, or
\begin{equation}
    \frac{R-P}{T-S}>\frac{1}{\sigma},
\end{equation}
where $\sigma$ is the structure coefficient independent of the payoff values. By substituting the condition (\ref{pointgrouptwogeneral}) we obtained in the donation game, we can determine the structure coefficient $\sigma$,
\begin{equation}
\sigma=\frac{(b/c)^*+1}{(b/c)^*-1}=\frac{1+p^{(1,1)}+w(1-p^{(1,1)})-\frac{4}{N}}{1-p^{(1,1)}+w(1+p^{(1,1)})}.
\end{equation}

When $p^{(1,1)}>1/(N-1)$, $1/\sigma$ increases with $w$, and cooperation is disfavored by increasing $w$. According to the rank $T>R>P>S$, we have $(R-P)/(T-S)<1$, which means cooperation is never favored if $1/\sigma>1$. To ensure $1/\sigma<1$, we have $p^{(1,1)}>2/(N(1-w))>2/N>1/(N-1)$. To sum up, in arbitrary two-player prisoner's dilemmas, cooperation is either disfavored as $w$ increases or unreachable at any $w$.

\section{Conclusion}\label{conclusion}
It is a frequently used assumption in evolutionary game dynamics that individuals prefer learning those strategies which provide higher fitness. However, driven by cognitive biases, one's fitness in perceptions when learning strategies could vary. In this work, we generalize the death-birth learning process with the consideration of a focal weight which makes it possible not to completely ignore the focal player's status in the death-birth protocol. More precisely, a higher weight provides an extra significance to the fitness of a focal player, which can reduce the frequency of changing strategies for both cooperation and defection. Despite the strategy-neutral character of this extension, we found that the usage of focal weight actually favors defection and hinders cooperation during the evolution. This is supported by the fact that the threshold $(b/c)^*$ of cooperation success increases as we enlarge the focal weight $w$.

Our theoretical analysis revealed a non-trivial tipping point of weight factor $w^\star=1-2k/N$, over which $(b/c)^*$ flips from positive to negative infinity and cooperation becomes unreachable. Importantly, such a tipping point always exists in a finite population when $k<N/2$. To find a simple testable example, we considered a square lattice topology with periodic boundary conditions where MC simulations can be executed. Our numerical calculations confirmed the theoretical predictions and underlined the validity of the results.

Furthermore, we verified the robustness of our observations from two perspectives. First, our conclusions remain valid for dispersal graphs with arbitrary edge weights and when the interaction and dispersal graphs overlap in an arbitrary way. Second, the conclusions do not change in arbitrary two-player prisoner's dilemmas.

Last, our results, valid in the weak selection limit, contradict those observations obtained for spatial populations under intermediate or strong selection strength. It was generally reported that if we introduce a sort of behavioral inertia, which helps players to maintain their strategies longer, then such modification of the microscopic dynamics can support cooperation significantly \cite{szolnoki_pre09,liu_yk_pa13,szolnoki2018reciprocity,zhang_yl_pre11,chen_xj_ijmpc08,jia_dy_pa18,szolnoki_csf20,chang_sh_pa18}. This phenomenon can be explained by the fact that cooperation and defection spread with significantly different speeds in a structured population. While cooperators advance slowly because they need to build a protective domain, defection can invade fast because it can enjoy the company of akin players. When we introduce inertia, propagation is slowed down for both cases, but in a biased way: defector propagation suffers more, resulting in a cooperator-supporting mechanism. These diverse conclusions provide an example when the evolutionary outcomes of the strong and weak selection limits are not comparable \cite{roca_plr09,fu_pre09b,li_c_pone13,zhong_wc_bs13}.

A.S. was supported by the National Research, Development and Innovation Office (NKFIH) under Grant No. K142948.

\end{document}